\documentclass[copyright]{eptcs}
\providecommand{\event}{PROLE 2014} 

\usepackage{breakurl}             
\usepackage[english]{babel}
\usepackage{verbatim}
\usepackage{alltt}
\usepackage{graphicx}
\usepackage{epstopdf}
\usepackage{multirow}
\usepackage{varwidth}

\newcommand{\etexttt}[1]{\textit{#1}}
\newcommand{\sepv}{\vspace{-3.0mm}}

\makeatletter
\g@addto@macro\@verbatim\footnotesize
\makeatother

\title{The ModelCC Model-Driven Parser Generator}
\author{Fernando Berzal \quad Francisco J. Cortijo \quad Juan-Carlos Cubero \quad Luis Quesada
\institute{CITIC \& Department of Computer Science and Artificial Intelligence\\University of Granada, Spain}
\email{\{berzal$|$cb$|$jc.cubero$|$lquesada\}@modelcc.org}
}
\def\titlerunning{The ModelCC Model-Driven Parser Generator: A Tutorial}
\def\authorrunning{F. Berzal, F.J. Cortijo, J.C. Cubero \& L. Quesada}
\begin{document}
\maketitle


\begin{abstract}
Syntax-directed translation tools require the specification of a language by means of a formal grammar. This grammar must conform to the specific requirements of the parser generator to be used. This grammar is then annotated with semantic actions for the resulting system to perform its desired function. In this paper, we introduce ModelCC, a model-based parser generator that decouples language specification from language processing, avoiding some of the problems caused by grammar-driven parser generators. ModelCC receives a conceptual model as input, along with constraints that annotate it. It is then able to create a parser for the desired textual syntax and the generated parser fully automates the instantiation of the language conceptual model. ModelCC also includes a reference resolution mechanism so that ModelCC is able to instantiate abstract syntax graphs, rather than mere abstract syntax trees.
\end{abstract}


\section{Introduction}

Widely-used language processing tools 
require language designers to provide a textual description of the language syntax, typically using a BNF-like notation. The proper specification of such a grammar is a nontrivial process that depends on the lexical and syntactic analysis techniques to be used, since particular techniques require the grammar to comply with specific and different constraints.
The most significant constraints on formal language specification originate from the need to consider context-sensitivity, the need of performing an efficient analysis, and some techniques' inability to consider grammar ambiguities or resolve conflicts caused by them.

Whenever the language syntax has to be modified, the language designer has to manually propagate the changes throughout the entire language processor tool chain. These updates are time-consuming, tedious, and error-prone. By making such changes labor-intensive, the traditional approach hampers the maintainability and evolution of the language \cite{Kats2010}.

Moreover, it is not uncommon that different tools use the same language, e.g. compilers, code generators, debuggers, lint-like utilities, code beautifiers... Multiple copies of the same language specification must then be maintained in sync, since language specification (i.e. its grammar) is tightly coupled to language processing (i.e. the semantic actions that annotate that grammar).

A grammar is a model of the language it defines. But a language can also be defined by a conceptual data model that represents the abstract syntax of the desired language, focusing on the elements the language will represent and their relationships. In conjunction with the declarative specification of some constraints, such model can be automatically converted into a grammar-based language specification. 

By using an annotated conceptual model, model-based language specification completely decouples language specification from language processing, which can be performed using whichever parsing techniques might be suitable for the formal language implicitly defined by the model. Semantic actions are no longer embedded within the language specification, as usual in grammar-driven language processors. The model representing the language can be modified as needed, without having to worry about the language processor and the peculiarities of the chosen parsing technique, since the corresponding language processor will be automatically updated. As the language model is not bound to any particular parsing technique, evaluating alternative and/or complementary parsing techniques is therefore possible without having to propagate their constraints into the language model.

It should be noted that, while the result of the traditional parsing process is an abstract syntax tree that corresponds to a valid interpretation of the input text according to the language syntax, nothing prevents the model-based language designer from modeling non-tree structures.
Indeed, a model-driven parser generator can automate the implementation of reference resolution mechanisms, among other syntactic and semantic checks that are typically deferred to later stages in the traditional language processing pipeline. 
Since ModelCC is able to resolve references, it obtains abstract syntax graphs as the result of the parsing process, rather than the abstract syntax trees obtained from conventional parser generators.

\section{Model-Based Language Specification} \label{sec:modelbased}

In this Section, we introduce the distinction between abstract and concrete syntax (\ref{subsec:asmcsm}), discuss the potential advantages of model-based language specification (\ref{subsec:modelbased}), and compare our approach with the traditional grammar-driven language design process (\ref{subsec:comparison}).

\subsection{Abstract Syntax and Concrete Syntaxes} \label{subsec:asmcsm}

The abstract syntax of a language is just a representation of the structure of the different elements of the language without the superfluous details related to its particular textual representation \cite{Kleppe2007}.
A concrete syntax is a particularization of the abstract syntax that defines, with precision, a specific textual or graphical representation of the language.
It should be noted that a single abstract syntax can be shared by several concrete syntaxes \cite{Kleppe2007}.

For example, the abstract syntax of the typical \emph{$<$if$>$-$<$then$>$-$<$optional else$>$} statement in imperative programming languages could be described as the concatenation of a conditional expression and one or two statements.
Different concrete syntaxes could be defined for such an abstract syntax, which would correspond to different textual representations of a conditional statement, e.g. \{``{\tt if}'', ``{\tt (}'', expression, ``{\tt )}'', statement, optional ``{\tt else}'' followed by another statement\} and \{``{\tt if}'', expression, ``{\tt then}'', statement, optional ``{\tt else}'' followed by another statement, ``{\tt endif}''\}.

The idea behind model-based language specification is that, starting from a single abstract syntax model (ASM) representing the core concepts in a language, language designers would later develop one or several concrete syntax models (CSMs).
These concrete syntax models would suit the specific needs of the desired textual or graphical representation for the language sentences.
The ASM-CSM mapping could be performed, for instance, by annotating the abstract syntax model with the constraints needed to transform the elements in the abstract syntax into their concrete representation.

\subsection{Advantages of Model-Based Language Specification} \label{subsec:modelbased}

Focusing on the abstract syntax of a language offers some benefits \cite{Kleppe2007} and provides some potential advantages to model-based language specification over the traditional grammar-based language specification approach:

\begin{itemize}

\item
When reasoning about the features a language should include, specifying its abstract syntax seems to be a better starting point than working on its concrete syntax details.
We control complexity by building abstractions that hide details when appropriate. 

\item
Sometimes, different incarnations of the same abstract syntax might be better suited for different purposes: an human-friendly syntax for manual coding, a machine-oriented format for automatic code generation, a Fit-like 
syntax for testing, different architectural views for discussions with project stakeholders...
It might be useful for a given language to support multiple syntaxes.

\item
Since model-based language specification is independent from specific lexical and syntax analysis techniques, the constraints imposed by specific parsing algorithms do not affect the language design process.
In principle, it might not even be necessary for the language designer to have advanced knowledge on parser generators when following a model-driven approach.

\item
A full-blown model-driven language workbench would allow the modification of a language abstract syntax model and the automatic generation of a working IDE on the run.
The specification of domain-specific languages would become easier, as the language designer could play with the language specification and obtain a fully-functioning language processor on the fly, without having to worry about the propagation of changes throughout the complete language processor tool chain.
\end{itemize}

In summary, the model-driven language specification approach brings domain-driven design \cite{ddd} to the domain of language design.
It provides the necessary infrastructure for what Evans would call the `supple design' of language processing tools: the intention-revealing specification of languages by means of abstract syntax models, the separation of concerns in the design of language processing tools by means of declarative ASM-CSM mappings, and the automation of a significant part of the language processor implementation.

\subsection{Comparison with the Traditional Approach} \label{subsec:comparison}

\begin{figure*}[tb!]
\centering
\begin{minipage}[t]{.5\textwidth}
  \centering
  \includegraphics[scale=0.2]{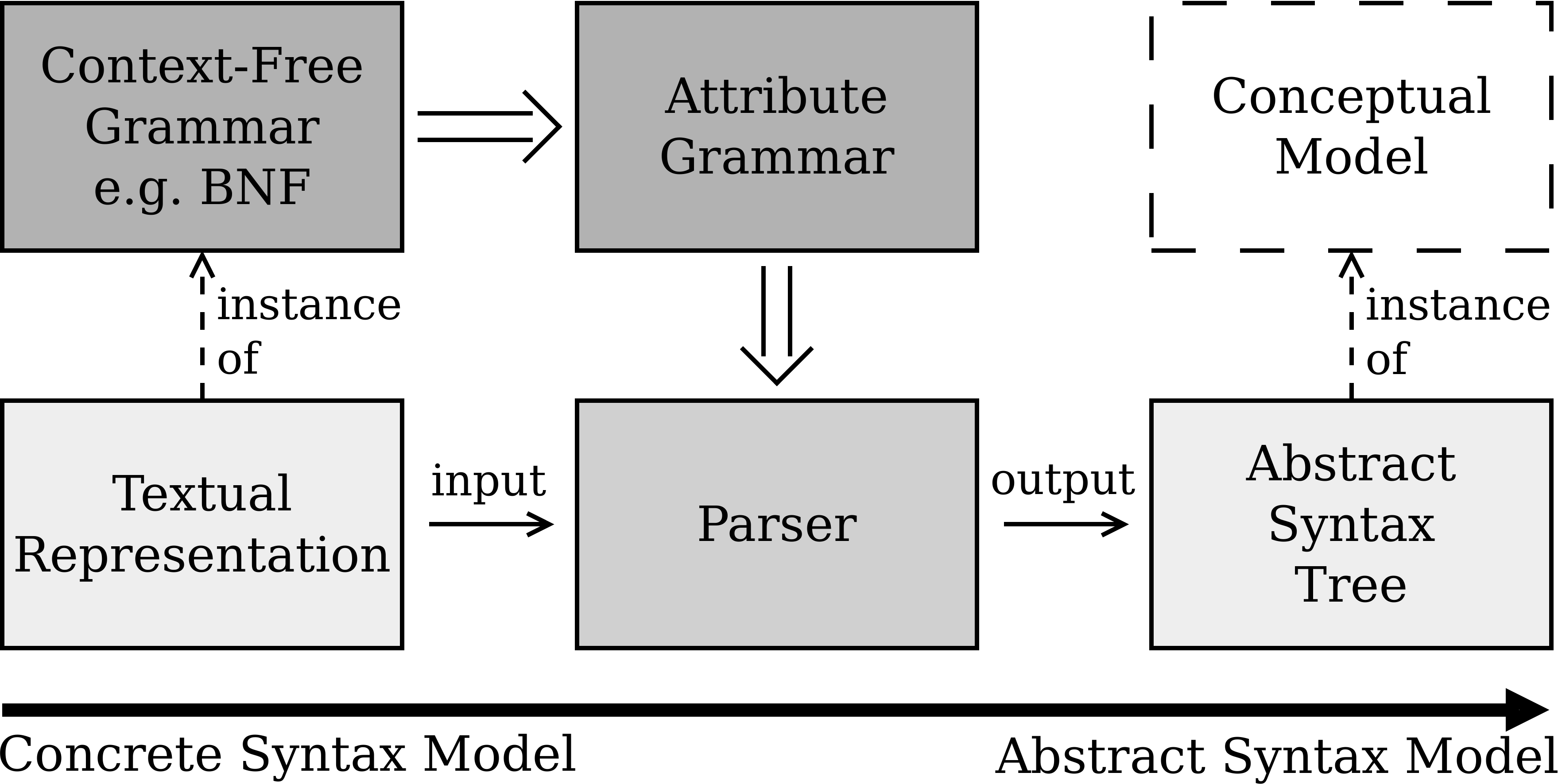}
  \label{fig:traditional}
\end{minipage}%
\begin{minipage}[t]{.5\textwidth}
  \centering
  \includegraphics[scale=0.2]{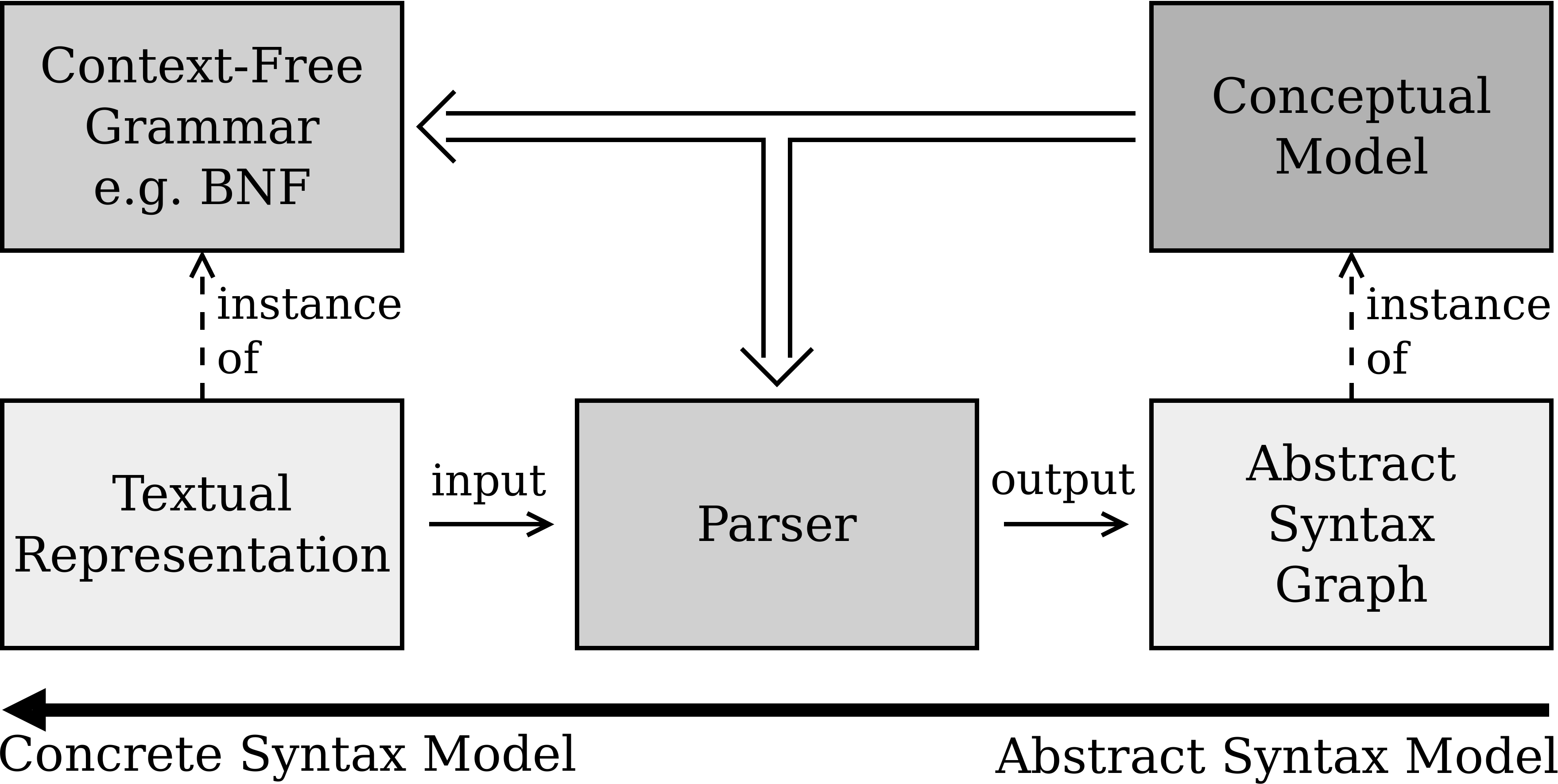}
  \label{fig:ModelCC}
\end{minipage}
\caption{The traditional (left) and the model-driven (right) language specification approaches.}
\label{fig:approach}
\end{figure*}

A diagram contrasting two different approaches to language specification and design is shown in Figure \ref{fig:approach}: the traditional grammar-driven approach on the left and the model-driven approach on the right.

When following the traditional grammar-driven approach, the language designer starts by designing the grammar corresponding to the concrete syntax of the desired language, typically in BNF or a similar notation.
Then, the designer annotates the grammar with attributes (and, probably, semantic actions), so that the resulting attribute grammar can be fed into lexer and parser generator tools that produce the corresponding lexers and parsers.
The syntax-directed translation process generates abstract syntax trees from the textual representation in the concrete syntax of the language.

When following the model-driven approach, the language designer starts by designing the conceptual model that represents the abstract syntax of the desired language, focusing on the elements the language will represent and their relationships.
Instead of dealing with the syntactic details of the language from the start, the designer devises a conceptual model for it (i.e. the abstract syntax model, or ASM), the same way a database designer starts with an implementation-independent conceptual database schema before he converts that schema into a logical schema that can be implemented in the particular kind of DBMS that will host the resulting database.
In the model-driven language design process, the ASM would play the role of entity-relationship diagrams in database design and each particular CSM would correspond to final table layouts for the physical database schema in relational DBMS's.

Even though the abstract syntax model of the language could be converted into a suitable concrete syntax model automatically, the language designer will often be interested in specifying the details of this ASM-CSM mapping.
With the help of constraints imposed over the abstract model, the designer is able to guide the conversion from the ASM to its concrete representation using a particular CSM.
This concrete model, when it corresponds to a textual representation of the abstract model, can be be described by a formal grammar.
It should be noted, however, that the specification of the ASM is independent from the peculiarities of the desired CSM.
Therefore, the grammar specification constraints enforced by particular parsing tools do not impose limits on the design of the ASM.
The model-driven language processing tool will take charge of those constraints and derive the grammar resulting from the ASM-CSM mapping that satisfies the parsing tool requirements.

While the traditional language designer specifies the grammar for the concrete syntax of the language, annotates it for syntax-directed processing, and obtains an abstract syntax tree that is an instance of the implicit conceptual model defined by the grammar, the model-based language designer starts with an explicit full-fledged conceptual model and specifies the necessary constraints for the ASM-CSM mapping.
In both cases, parser generators create the tools that parse the input text in its concrete syntax.
The difference lies in the specification of the grammar that drives the parsing process, which is hand-crafted in the traditional approach and automatically-generated as a result of the ASM-CSM mapping in the model-driven approach.

Another difference stems from the fact that the result of the parsing process is an instance of an implicit model in the grammar-driven approach while that model is explicit in the model-driven approach.
An explicit conceptual model is absent in the traditional language design process albeit that does not mean that it does not exist.
The model-driven approach enforces the existence of an explicit conceptual model, which lets the proposed approach reap the benefits of domain-driven design.

In general, the result of the parsing process is an abstract syntax tree that corresponds to a valid interpretation of the input text according to the language concrete syntax (at least for the constituency-based parsers typically used for programming languages). However, nothing prevents the conceptual model designer from modeling non-tree structures, which can be described, for instance, by reference attributed grammars \cite{Burger2010}.
Hence the use of the `abstract syntax graph' term in Figure \ref{fig:ModelCC}.
This might be useful, for instance, for modeling graphical languages, which are not constrained by the linear nature of the text-based languages.

In summary, the model-based language specification process goes from the abstract to the concrete, instead of following the traditional syntax-directed approach that goes from a concrete syntax model to an implicit abstract syntax model, which is now explicit in the model-driven approach. This alternative approach facilitates the proper design and implementation of language processing systems by decoupling language processing from language specification.

\section{ModelCC Model Specification} \label{sec:modelspecification}

Once we have described model-driven language specification in general terms, we now proceed to introduce ModelCC \cite{Quesada2014c}, a tool that supports the model-driven approach for thee design of language processing systems. ModelCC, at its core, acts as a parser generator. The starting abstract syntax model is created by defining classes that represent language elements and establishing relationships among those elements (associations in UML terms). Once the abstract syntax model is established, its incarnation as a concrete syntax is guided by the constraints imposed over language elements and their relationships as annotations on the abstract syntax model. In other words, the declarative specification of constraints over the ASM establishes the desired ASM-CSM mapping.

\begin{figure*}[t!]
\centering
\includegraphics[scale=0.6]{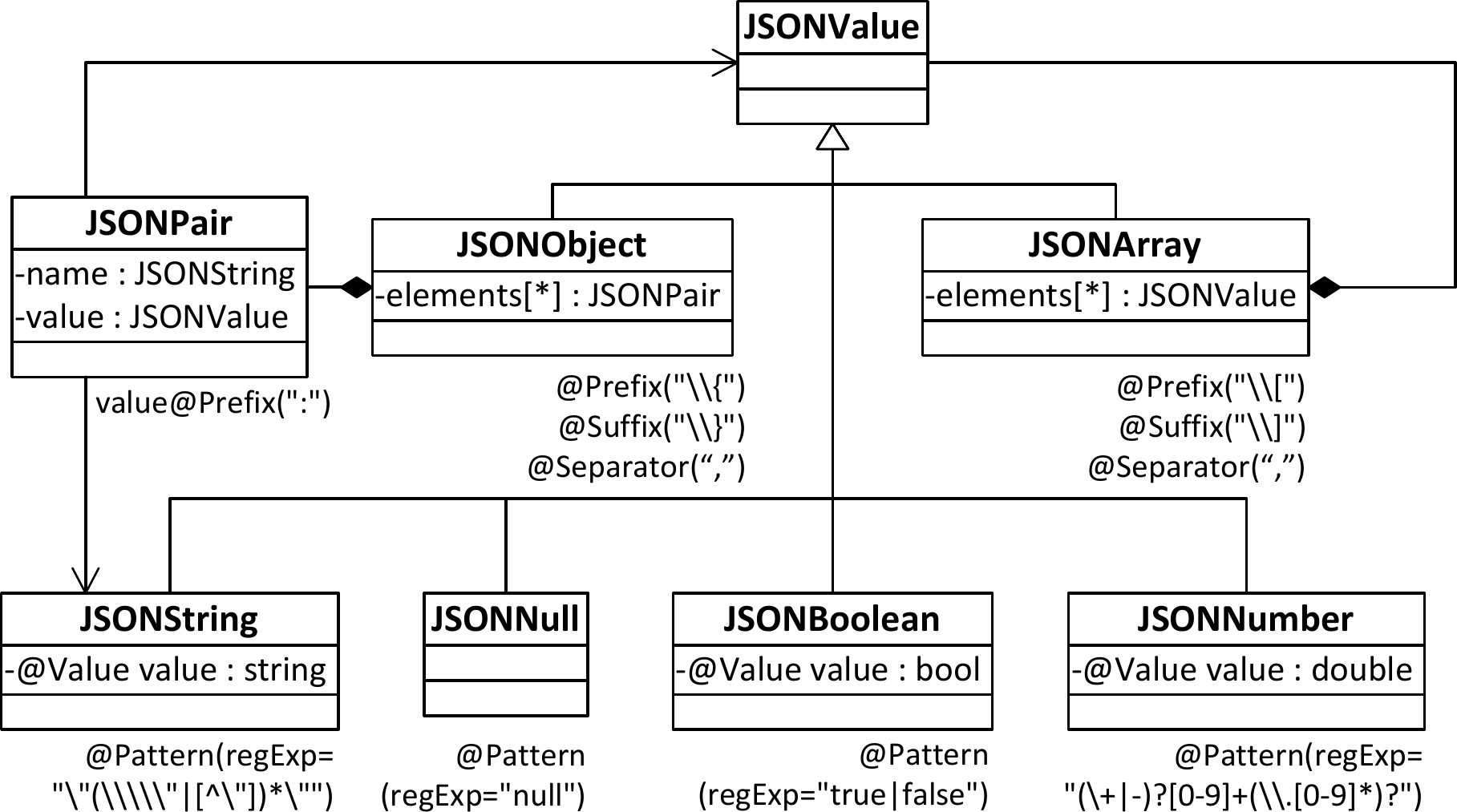}
\caption{ModelCC specification of the JSON open standard for data exchange.} \label{fig:language-json}
\end{figure*}

ModelCC allows the specification of languages in the form of abstract syntax models such as the one shown in Figure \ref{fig:language-json}.
This model, depicted here as an UML class diagram for clarity, specifies the abstract syntax model of the JSON open data exchange format.
In Section \ref{sec:example1}, we will analyze a more traditional example: the language of arithmetic expressions (see Figure \ref{fig:calcmodelcc}).
The annotations that accompany the model in Figure \ref{fig:language-json} provide the necessary information for establishing the complete ASM-CSM mapping that results in the concrete syntax of the JSON standard (as the annotations in Figure \ref{fig:calcmodelcc} define the traditional infix notation for arithmetic expressions).

In this Section, we introduce the basic constructs that allow the specification of abstract syntax models, while we will discuss how model constraints help us establish the desired ASM-CSM mapping in Section \ref{sec:modelconstraints}.
Basically, the ASM is built on top of basic language elements, which might be viewed as the tokens of the model-driven language specification.
Model-driven language processing tools such as ModelCC provide the necessary mechanisms to combine those basic elements into more complex language constructs, which correspond to the use of concatenation, selection, and repetition in the syntax-driven specification of languages.


\subsection{Concatenation}


Concatenation is the most basic construct we can use to combine sets of language elements into more complex language elements.
In textual languages, this is achieved just by joining the strings representing its constituent language elements into a longer string that represents the composite language element.
In ModelCC, concatenation is achieved by object composition.
The resulting language element is the composite element and its members are the language elements the composite element collates.

In Figure \ref{fig:language-json}, {\tt JSONPair} is a composite element that results from the concatenation of a {\tt JSONString} name and a {\tt JSONValue} value.

When translating the ASM into a textual CSM, each composite element in a ModelCC model generates a production rule in the grammar representing the CSM.
This production, with the nonterminal symbol of the composite element in its left-hand side, concatenates the nonterminal symbols corresponding to the constituent elements of the composite element in its right-hand side.
By default, the order of the constituent elements in the production rule is given by the order in which they are specified in the model, but such an order is not mandatory (e.g. many ambiguous languages require different ordered sequences of constituent elements and even some unambiguous languages allow for unordered sequences of constituent elements).


\subsection{Selection}

Selection is the language modeling construct used to represent choices: it enables alternative elements in language constructs. In ModelCC, selection is created by subtyping. Specifying object-oriented inheritance relationships between language elements is equivalent to defining `is-a' relationships in  traditional database design. The language element we wish to establish alternatives for is the superelement (i.e. the superclass in OO design or the supertype in DB modeling), whereas the different alternatives are represented as subelements (i.e. subclasses in OO, subtypes in DB modeling).
Alternative elements are always kept separate to enhance the modularity of ModelCC abstract syntax models and their integration in language processing systems.

In Figure \ref{fig:language-json}, {\tt JSONValue}s can be either {\tt JSONObject}s or {\tt JSONArray}s, as well as a bunch of basic data values including strings ({\tt JSONString}), numbers ({\tt JSONNumber}), booleans ({\tt JSONBoolean}), and null ({\tt JSONNull}).



Each inheritance relationship in ModelCC, when converting the ASM into a textual CSM, generates a production rule in the CSM grammar.
In those productions, the nonterminal symbol corresponding to the superelement appears in its left-hand side, while the nonterminal symbol of the subelement appears as the only symbol in the production right-hand side.
Obviously, if a given superelement has $k$ different subelements, $k$ different productions will be generated representing the $k$ alternatives defined in the ASM.


\subsection{Repetition}

Representing repetition is also necessary in abstract syntax models, since a language element might appear several times in a given language construct, but, when a variable number of repetitions is allowed, mere concatenation does not suffice to model it in the ASM. Repetition is also achieved through object composition in ModelCC, just by allowing different multiplicities in the associations that connect composite elements to their constituent elements.

In Figure \ref{fig:language-json}, {\tt JSONObject}s are made of a variable number of {\tt JSONPair}s. Likewise, {\tt JSONArray}s contain a variable number of {\tt JSONValue}s.

Each composition relationship representing a repetitive structure in the ASM will lead to two additional production rules in the grammar defining its textual CSM. A recursive production of the form \etexttt{$<$List$>$ ::= $<$Element$>$ $<$List$>$} allows for the repetition of elements, whereas a simple production  \etexttt{$<$List$>$ ::= $\epsilon$} or \etexttt{$<$List$>$ ::= $<$Element$>$} provides the base case for the recursion, depending on whether the list can be empty or not.
It should also be noted that \etexttt{$<$List$>$} will take the place of the \etexttt{$<$Element$>$} nonterminal in the production derived from the composition relationship that connects the repeating element with its composite element. Element multiplicities and list delimiters in the CSM will be determined from the constraints we will now see in Section \ref{sec:modelconstraints}

%
%
%

\section{ModelCC Model Constraints} \label{sec:modelconstraints}

\begin{table*}[tb!]
\begin{center}

\setlength{\tabcolsep}{7pt}
\resizebox{\linewidth}{!}{
\begin{tabular}{ l  l  l } \hline
\normalsize
Constraints on & Annotation & Function \\ \hline

\multirow{2}{*}{... patterns}
& @Pattern & Pattern matching specification of basic language elements. \\
& @Value & Field where the recognized input token will be stored. \\ \hline

\multirow{3}{*}{... delimiters}
& @Prefix & Element prefix(es). \\
& @Suffix & Element suffix(es). \\
& @Separator & Element separator(s) in lists of elements. \\ \hline

\multirow{2}{*}{... cardinality}
& @Optional & Optional elements.\\
& @Multiplicity & Minimum and maximum element multiplicity.\\ \hline

\multirow{3}{*}{... evaluation order}
& @Associativity & Element associativity (e.g. left-to-right). \\
& @Composition & Eager or lazy composition for nested composites. \\
& @Priority & Element precedence level/relationships. \\ \hline

\multirow{2}{*}{... composition order}
& @Position & Element member relative position. \\
& @FreeOrder & When there is no predefined order among element members. \\ \hline

\multirow{2}{*}{... references}
& @ID & Identifier of a language element. \\
& @Reference & Reference to a language element. \\ \hline

\multirow{2}{*}{Custom constraints}
& \multirow{2}{*}{@Constraint} & \multirow{2}{*}{Custom user-defined constraint.} \\ \\ \hline
\end{tabular}
}
\end{center}
\caption{The constraints supported by the ModelCC model-based parser generator.} \label{fig:tablesummary}
\end{table*}

Once we have examined the mechanisms that let us create abstract syntax models in ModelCC, we now proceed to describe how constraints can be imposed on such models in order to establish the desired ASM-CSM mapping.

\begin{figure*}[t!]
\centering
\centering
\begin{minipage}[t]{.35\textwidth}
  \centering
  \includegraphics[scale=0.6]{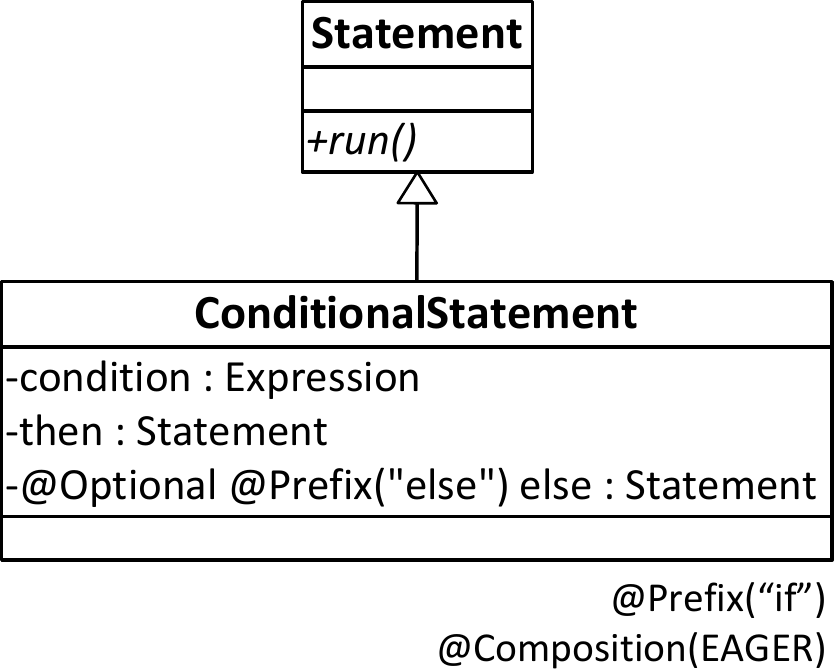}
  \label{fig:test1}
\end{minipage}%
\begin{minipage}[t]{.65\textwidth}
  \centering
  \includegraphics[scale=0.6]{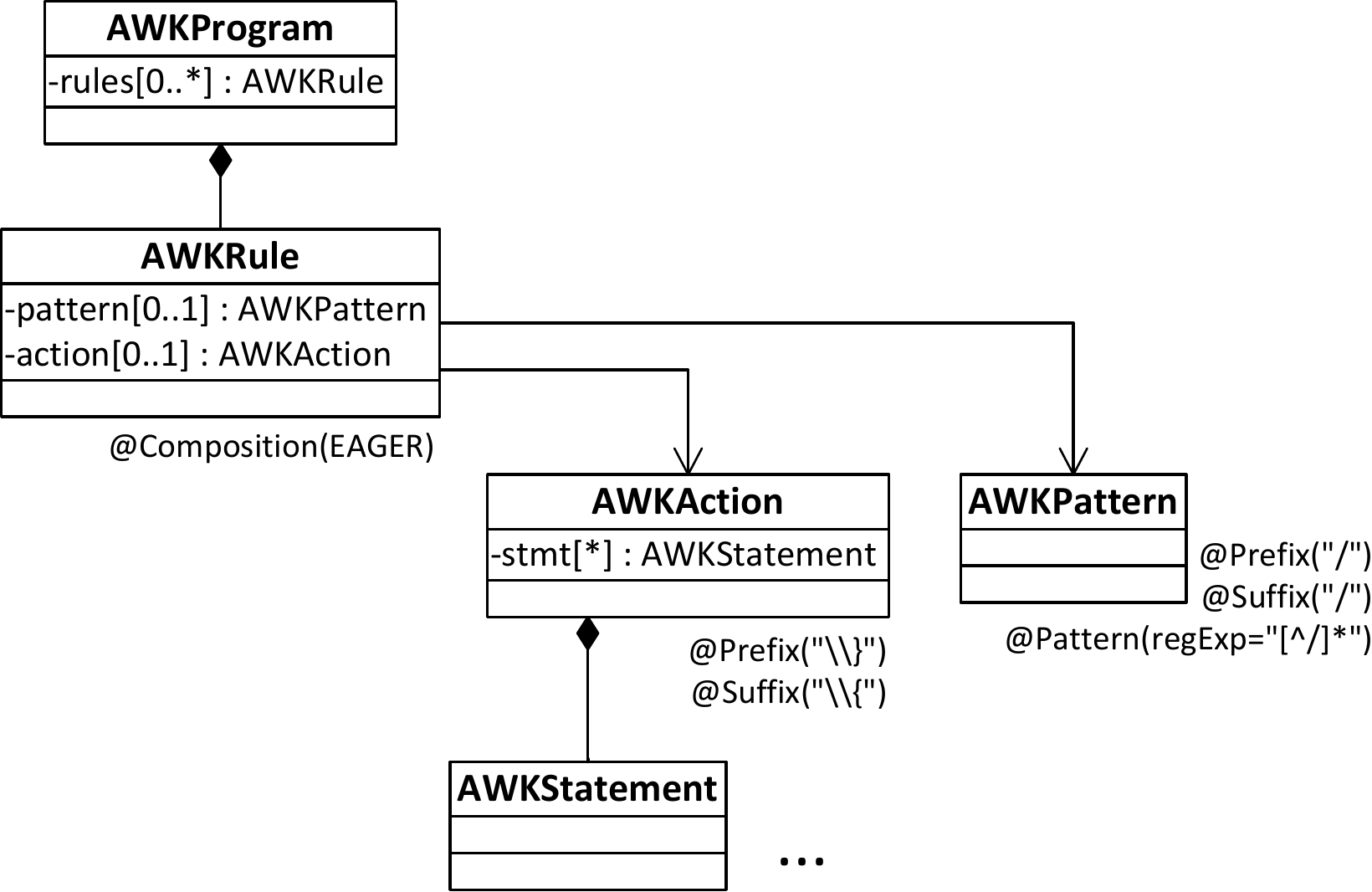}
  \label{fig:test2}
\end{minipage}
\caption{ModelCC specification and resolution of shift-reduce conflicts: composite if-then-else statements (left) and the AWK programming language (right). In nested conditional statements with optional else statements but without and end-if marker, the conflict is resolved by associating the dangling else with the nearest if (eager composition). In the AWK programming language, rules consist of actions and patterns. Since both actions and patterns can be missing, but not both at the same time, a similar shift-reduce conflict appears when we take into account that an AWK program is just a list of rules.} \label{fig:language-shift-reduce}
\end{figure*}

Table \ref{fig:tablesummary} summarizes the set of constraints supported by ModelCC for establishing the ASM-CSM mappings between abstract syntax models and their concrete representation in textual CSMs:

\begin{itemize}

\item
A first set of constraints is used for pattern specification, a necessary feature for defining the lexical elements of the concrete syntax model, i.e. its tokens. Pattern matching lets us extract fragments from the textual input (e.g. using regular expressions) and fill in values for the basic language elements that are the building blocks of more complex abstract syntax models.

\item
A second set of constraints is employed for defining delimiters and separators in the concrete syntax model. They help us eliminate language ambiguities, when we want to obtain deterministic context-free languages, or can be used just as syntactic sugar to help improve the readability and writability of many languages. They are also common in repeating elements, which can be annotated with separators (in case separators are employed, the recursive production derived from repeating elements will be of the form \etexttt{$<$List$>$ ::= $<$Element$>$ $<$Separator$>$ $<$List$>$}).

\item
A third set of ModelCC constraints lets us impose cardinality constraints on language elements, which can be used to control the multiplicity of repeating language elements, as well as the optionality or mandatoriness of any element in the language model.

\item
A fourth set of constraints lets us impose an evaluation order on language elements. These constraints are employed to declaratively resolve ambiguities in the concrete syntax of a textual language by establishing associativity, precedence, and composition policies. Associativity and precedence constraints are common in the evaluation of arithmetic expressions, as we will see in the next section (see, e.g., Figure \ref{fig:calcmodelcc}). Composition constraints help us resolve the ambiguities that cause the typical shift-reduce conflicts in LR parsers, as shown in Figure \ref{fig:language-shift-reduce}.

\item
A fifth set of constraints lets us specify the relative ordering of the constituents in composite language elements, or even allow for the free ordering of elements (a feature uncommon in programming languages, yet useful in other settings beyond deterministic context-free languages).

\item
A sixth set of constraints lets us specify referenceable language elements and references to them, enabling the reference resolution mechanism included in ModelCC. When references are resolved, the ModelCC parser returns an abstract syntax graph instead of the abstract syntax tree resulting from context-free grammar parsing.

\item
Finally, custom constraints let us provide specific lexical, syntactic, and semantic constraints that take into consideration additional context information. Certainly not needed for deterministic context-free grammars, they provide a general customization mechanism for ModelCC extensions.

\end{itemize}

As soon as the complete ASM-CSM mapping is established, ModelCC is able to generate the suitable parser for the concrete syntax defined by the CSM. In its current version, this ASM-CSM mapping is specified with the help of metadata annotations on the class model that defines the ASM. Now supported by all the major programming platforms, metadata annotations have been used in reflective programming and code generation \cite{Fowler2002}. Among many other things, they can be employed for dynamically extending the features of your software development runtime \cite{Berzal2005} or even for building complete model-driven software development tools that benefit from the infrastructure provided by your standard compiler and its associated tools \cite{mdsd-ideal}.

\section{A Simple Example} \label{sec:example1}

An interpreter for arithmetic expressions in infix notation can be used to illustrate the differences between ModelCC and more conventional tools. A full implementation of an extended example using ModelCC and two well-known parser generators (lex \& yacc on the one side, ANTLR on the other) is available at \url{http://www.modelcc.org/examples}. Albeit the arithmetic expression example is necessarily simplistic, it already provides some hints on the potential benefits that model-driven language specification can bring to more challenging endeavors. This simple language is also used in the next section as the basis for a more complex language, which illustrates ModelCC reference resolution mechanism.


Using conventional tools, the language designer would start by specifying the grammar defining the arithmetic expression language in a BNF-like notation.
When using lex \& yacc, the language designer converts the BNF grammar into a grammar suitable for LR parsing.
Likewise, when using ANTLR, the language designer converts the BNF grammar into a grammar suitable for LL parsing. LL(*) parsers do not support left-recursion, so left-recursive grammar productions must be refactored.
Since ANTLR provides no mechanism for the declarative specification of token precedences, such precedences must be incorporated into the grammar.
Unfortunately, these grammar refactorings typically involve the introduction of a certain degree of duplication in the language specification, such as separate token types in the lexer and multiple parallel production rules in the parser.
Once the grammar is adjusted to satisfy the constraints imposed by the parser generators, the language designer can define the semantic actions needed to implement our arithmetic expression interpreter.
Using lex \& yacc, albeit somewhat verbose using the C programming language syntax, the implementation of an arithmetic expression interpreter is relatively straightforward.
The streamlined syntax of the scannerless ANTLR parser generator makes this implementation significantly more concise than the equivalent lex \& yacc implementation.

\begin{figure*}[tb!]
\centering
\includegraphics[scale=0.6]{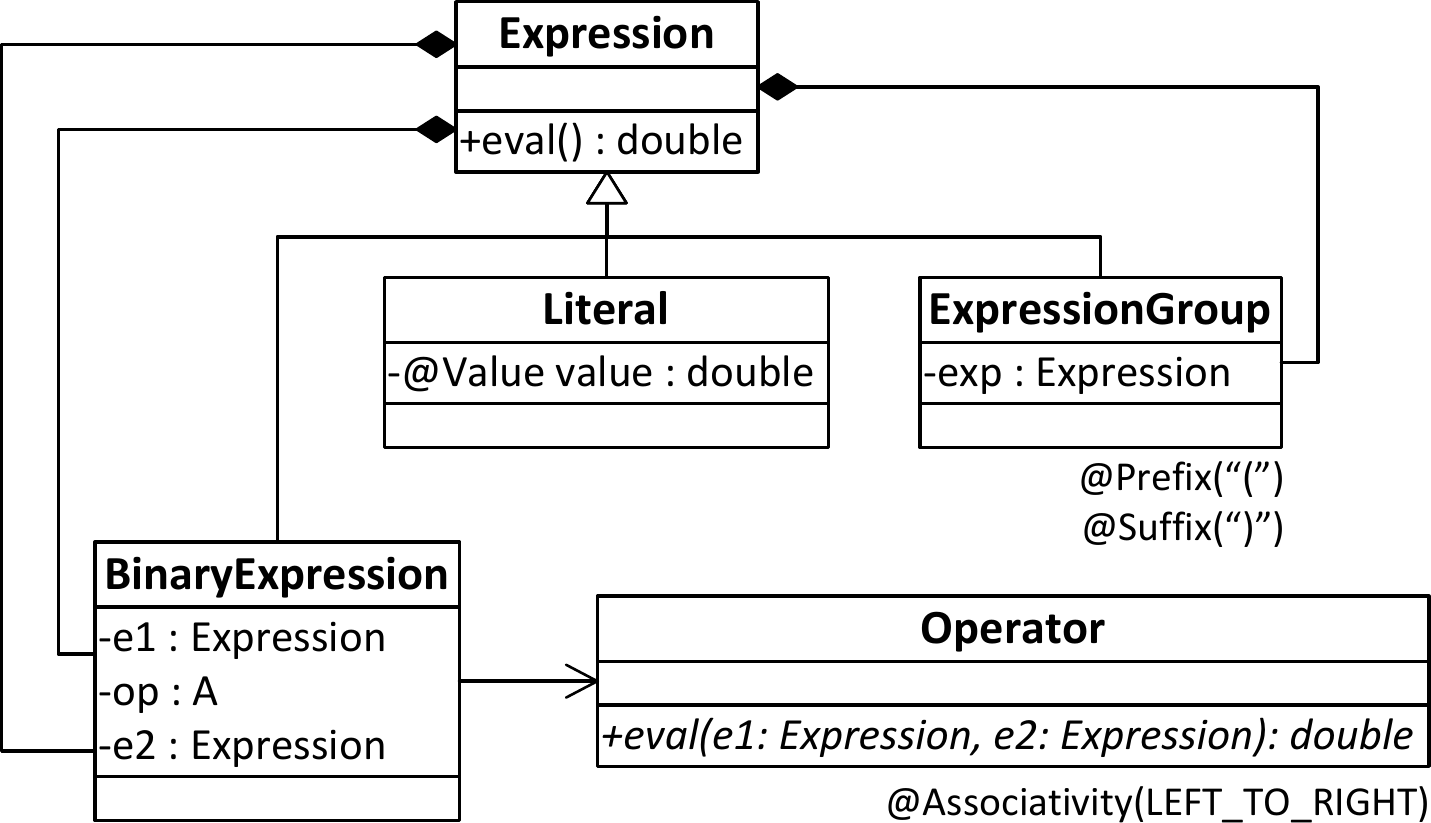}
\caption{ModelCC specification of the arithmetic expression language.} \label{fig:calcmodelcc}
\end{figure*}

When following a model-based language specification approach, the language designer starts by elaborating an abstract syntax model, which will later be mapped to a concrete syntax model by imposing constraints on the abstract syntax model.
Annotated models can be represented graphically, as the UML class diagram in Figure \ref{fig:calcmodelcc}, or implemented using conventional programming languages, as the complete Java implementation included in Figure \ref{fig:calcimmodelcc}. The declarative specification of associativity and precedence constraints for the different operators spare us from the grammar refactorings needed by conventional tools. The implementation of the arithmetic expression interpreter is also more elegant in ModelCC: the polymorphic {\tt eval()} method takes care of the evaluation of arithmetic expressions.

\begin{figure*}[tb!]
\begin{verbatim}
// Expressions
\end{verbatim}
\sepv
\begin{verbatim}
public abstract class Expression implements IModel {
  public abstract double eval();
}
\end{verbatim}
\sepv
\begin{verbatim}
@Prefix("\\(") @Suffix("\\)")
public class ExpressionGroup extends Expression {
  Expression e;
  @Override public double eval() { return e.eval(); }
}
\end{verbatim}
\sepv
\begin{verbatim}
public class Literal extends Expression {
  @Value double value;
  @Override public double eval() { return value; }
}
\end{verbatim}
\sepv
\begin{verbatim}
public class BinaryExpression extends Expression {
  Expression e1;
  Operator op;
  Expression e2;
  @Override public double eval() { return op.eval(e1,e2); }
}
\end{verbatim}
\sepv
\begin{verbatim}
// Operators
\end{verbatim}
\sepv
\begin{verbatim}
@Associativity(AssociativityType.LEFT_TO_RIGHT)
public abstract class Operator implements IModel {
  public abstract double eval(Expression e1,Expression e2);
}
\end{verbatim}
\sepv
\begin{verbatim}
@Pattern(regExp="\\+") @Priority(value=2)
public class AdditionOperator extends Operator {
  @Override public double eval(Expression e1,Expression e2) { return e1.eval()+e2.eval(); }
}
\end{verbatim}
\sepv
\begin{verbatim}
@Pattern(regExp="-") @Priority(value=2)
public class SubtractionOperator extends Operator {
  @Override public double eval(Expression e1,Expression e2) { return e1.eval()-e2.eval(); }
}
\end{verbatim}
\sepv
\begin{verbatim}
@Pattern(regExp="\\*") @Priority(value=1)
public class MultiplicationOperator extends Operator {
  @Override public double eval(Expression e1,Expression e2) { return e1.eval()*e2.eval(); }
}
\end{verbatim}
\sepv
\begin{verbatim}
@Pattern(regExp="\\/") @Priority(value=1)
public class DivisionOperator extends Operator {
  @Override public double eval(Expression e1,Expression e2) { return e1.eval()/e2.eval(); }
}
\end{verbatim}
\caption{Complete Java implementation of the arithmetic expression interpreter using ModelCC: Java classes define the language ASM,
metadata annotations specify the desired ASM-CSM mapping, and the {\tt eval()} method implements arithmetic expression evaluation.}
\label{fig:calcimmodelcc}
\end{figure*}

Figure \ref{fig:run} shows the actual code needed to generate and invoke the parser in ModelCC.
ModelCC generates a parser from the arithmetic expression language model. This parser receives input strings such as ``10/(2+3)*0.5+1'' and instantiates \emph{Expression} objects from them. The {\tt eval()} method then yields the final result of the evaluation (2 in this case).

\begin{figure*}[tb!]
\centering
\begin{verbatim}
// Read the model.
Model model = JavaModelReader.read(Expression.class);

// Generate the parser.
Parser<Expression> parser = ParserFactory.create(model);

// Parse the input string and instantiate the corresponding expression.
Expression expr = parser.parse("10/(2+3)*0.5+1");

// Evaluate the expression.
double value = expr.eval();
\end{verbatim}
\caption{Code snippet showing how the arithmetic expression parser is generated and the resulting interpreter is invoked.} \label{fig:run}
\end{figure*}

In its current version, ModelCC generates Lamb lexers \cite{Quesada2011a} and Fence parsers \cite{Quesada2012f}, albeit traditional LL and LR parsers might also be generated whenever the ASM-CSM mapping constraints make LL and LR parsing feasible.

ModelCC also provides a testing framework that integrates well with existing IDEs and JUnit.
Since separate language elements are models themselves, it is possible to implement unit tests that focus on specific language elements and integration tests for the successive refinements of a language model, hereby enabling and supporting the incremental design of languages.

Since the abstract syntax model in ModelCC is not constrained by the vagaries of particular parsing algorithms, the language design process can be focused on its conceptual design, without the artificial introduction of design artifacts just to satisfy the demands of particular tools:

\begin{itemize}

\item
Conventional tools such as lex \& yacc example force the creation of artificial token types in order to avoid lexical ambiguities, which leads to duplicate grammar production rules and duplicate semantic actions in the language specification.
As in any other software development project, duplication hinders the evolution of languages and affects the maintainability of language processors.
In ModelCC, duplication in the language model does not have to be included to deal with lexical ambiguities: token type definitions do not have to be adjusted, duplicate syntactic production rules will not appear in the language model, and, as a consequence, semantic predicates do not have to be duplicated either.

\item
Established parser generators require modifications to the language grammar in order to comply with parsing constraints, let it be the elimination of left-recursion for LL parsers or the introduction of new nonterminals so that the desired precedence relationships are established.
In the model-driven language specification approach, the left-recursion problem disappears since it is something the underlying tool can easily deal with in a fully-automated way when an abstract syntax model is converted into a concrete syntax model.
Moreover, the declarative specification of constraints is orthogonal to the abstract syntax model that defines the language.
Those constraints fully determine the ASM-CSM mapping and, since ModelCC takes charge of everything in the conversion process, the language designer does not have to modify the abstract syntax model just because a given parser generator might prefer its input in a particular format.
This is the main benefit that results from raising your abstraction level in model-based language specification.

\item
When changes in the language specification are necessary, as it is often the case when a software system is successful, the traditional language designer will have to propagate changes throughout the entire language processing tool chain, often introducing significant changes and making profound restructurings in the production code base.
These changes can be time-consuming, quite tedious, and extremely error-prone.
In contrast, modifications are easier when a model-driven language specification approach is followed.
Any modifications in the language will affect either to the abstract syntax model, when a language is extended with new capabilities, or to the constraints that define the ASM-CSM mapping, whenever syntactic details change or new CSMs are devised for the same ASM.
In either case, the more time-consuming, tedious, and error-prone modifications are automated and the language designer can focus his efforts on the essence of the required changes rather than on their accidents.

\item
Traditional parser generators typically mix semantic actions with the syntactic details of the language specification.
This approach, which might be justified when performance is the top concern, might lead to poorly-designed hard-to-test systems.
Moreover, when different applications or tools employ the same language, any changes to the syntax of that language must be carefully replicated in all the applications and tools that use the language.
The maintenance of several versions of the same language specification in parallel might also lead to severe maintenance problems.
In contrast, the separation of concerns provided by ModelCC, which separates ASM and ASM-CSM mappings, promotes a more elegant design for language processing systems.
By decoupling language specification from language processing and providing an explicit conceptual model for the language, different applications and tools can now use the same language without duplicate language specifications.
A similar result could be hand-crafted using traditional parser generators (i.e. making their implicit conceptual model explicit and working on that explicit model), but ModelCC automates this part of the process.

\end{itemize}

In summary, while traditional language processing tools provide different mechanisms for resolving ambiguities and implementing language constraints, the solutions they provide typically interfere with the conceptual modeling of languages: relatively minor syntactic details might significantly affect the structure of the whole language specification.
Model-driven language specification, as exemplified by ModelCC, provides a cleaner separation of concerns: the abstract syntax model is kept separate from its incarnation in concrete syntax models, thereby separating the specification of abstractions in the ASM from the particularities of their textual representation in CSMs.

\section{Additional Examples} \label{sec:example2}

ModelCC is able to automatically generate a grammar from the ASM defined by the class model and the ASM-CSM mapping, which is specified as a set of metadata annotations on the class model. These annotations also provide a mechanism for reference resolution that allows the automatic instantiation of complete object graphs.



\begin{figure*}[tb!]
\centering
\begin{verbatim}
// Read the model.
Model model = JavaModelReader.read(Expression.class);

// Create the parser.
Parser<Expression> parser = ParserFactory.create(model);

// Define a constant
parser.add(new Constant("pi", 3.1415927));

// Use the predefined constant in JUnit tests for arithmetic expressions
assertEquals(3.1415927,   parser.parse("pi").eval(),   EPSILON);
assertEquals(2*3.1415927, parser.parse("2*pi").eval(), EPSILON);
\end{verbatim}
\caption{Code snippet showing ModelCC support for separate compilation using predefined model elements.} \label{fig:constant}
\end{figure*}

\begin{figure*}[tb!]
\centering
\includegraphics[scale=0.6]{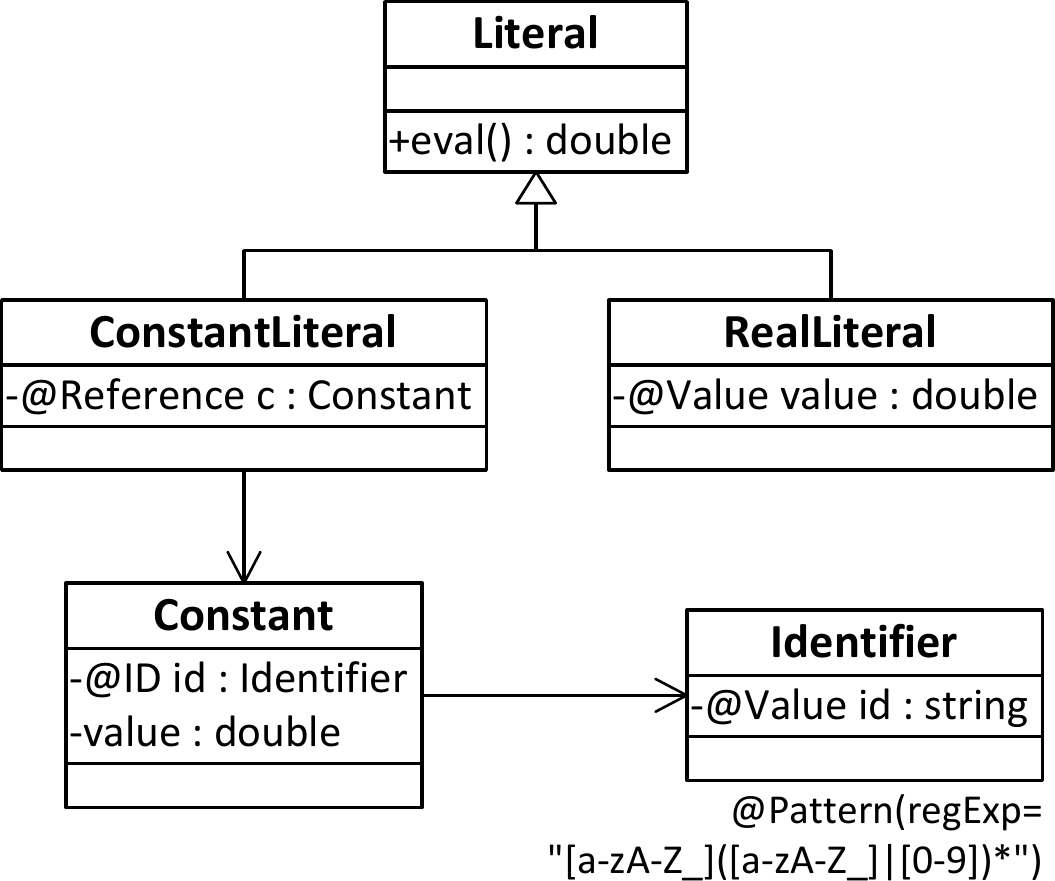}
\caption{ModelCC reference resolution mechanism.} \label{fig:language-reference}
\end{figure*}

The reference resolution mechanism in ModelCC is illustrated by the code snippet in Figure \ref{fig:constant} and the model shown in \ref{fig:language-reference}. In this example, which extends our arithmetic expression language, a constant is defined before the parser is invoked. Then we can parse an expression that includes references to the predefined constant, whose definition does not have to be included in the textual input of the parser, thus providing a crude but elegant form of separate compilation.

Following the same approach, we could easily design a full-fledged imperative programming language. Language composition would enable us to extend our arithmetic expression language easily, just by including statements, new expression types, and additional operators in our language model.

\begin{figure*}[tb]
\centering
\includegraphics[scale=0.6]{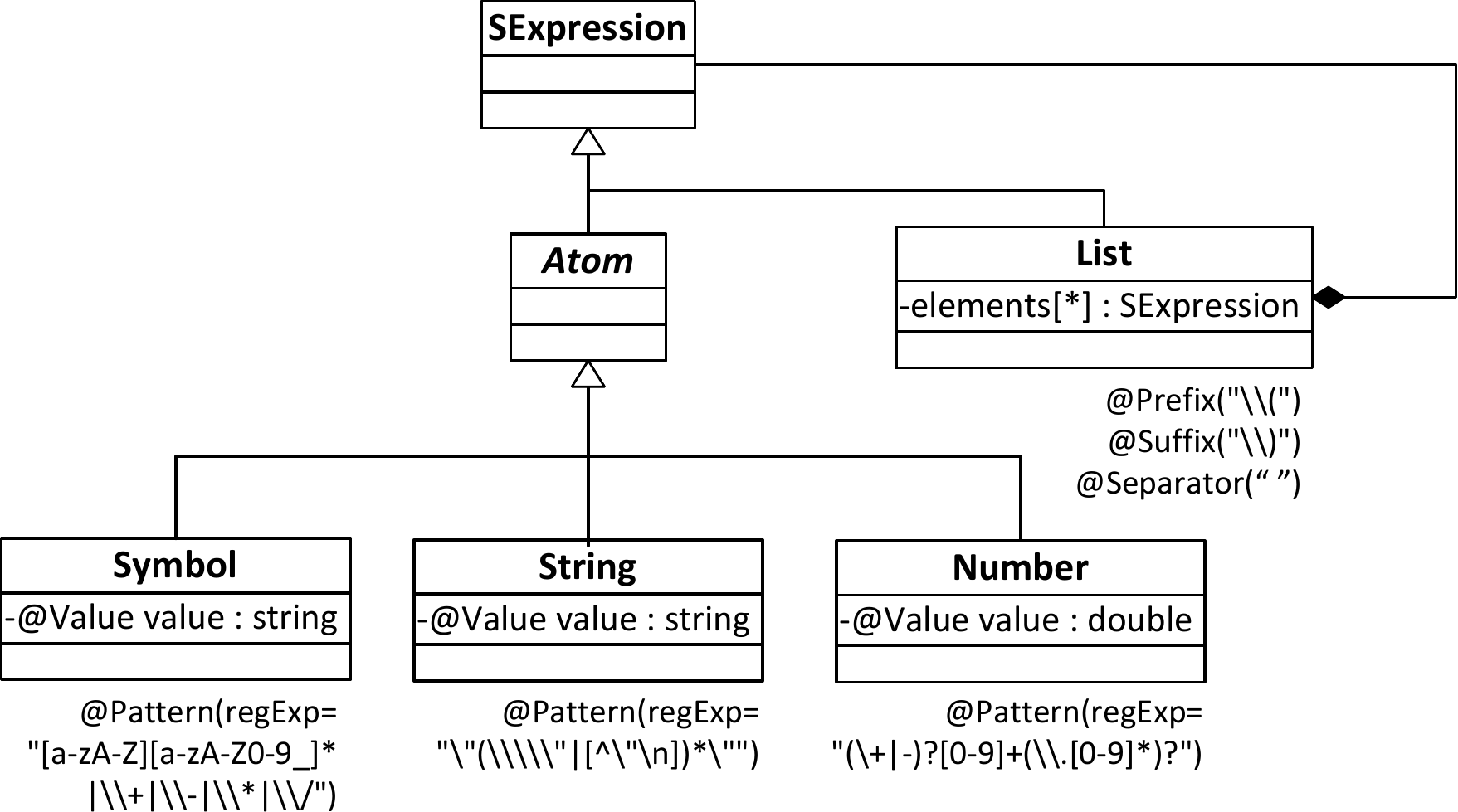}
\caption{ModelCC specification of the S-expression syntax of LISP-like functional languages.} \label{fig:language-lisp}
\end{figure*}

\begin{figure*}[tb]
\centering
\includegraphics[scale=0.6]{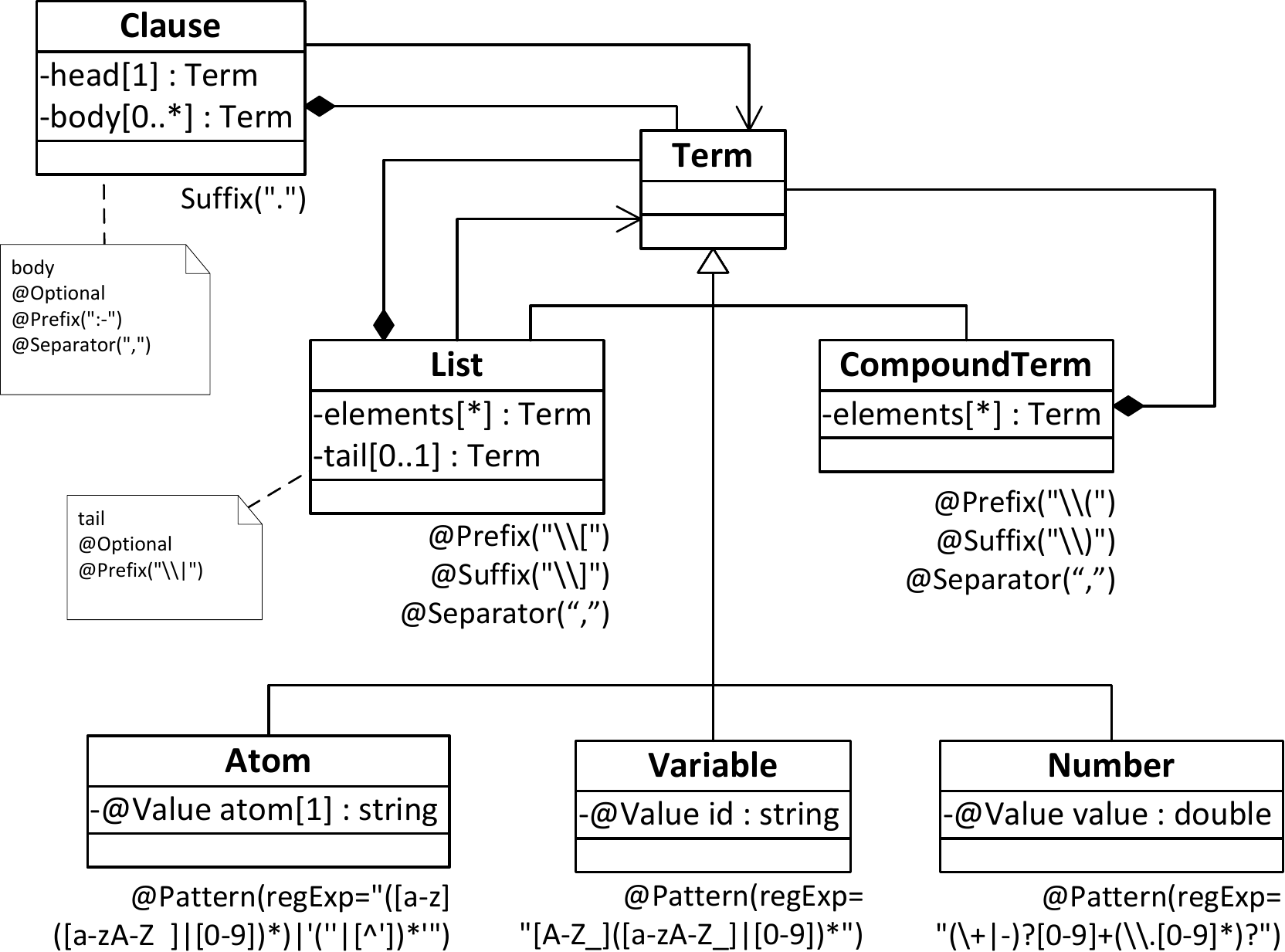}
\caption{ModelCC specification of the syntax of the PROLOG logic programming language.} \label{fig:language-prolog}
\end{figure*}

Figures \ref{fig:language-lisp} and \ref{fig:language-prolog} include two more examples: the traditional syntax of LISP S-expressions and a PROLOG-like logic programming languages. A fully-functional version of ModelCC for Java, additional examples of its use, and a detailed user manual describing all the annotations that can be used to annotate class models in ModelCC can be found at the ModelCC web site: \url{http://www.modelcc.org}.

\section{Conclusions and Future Work} \label{sec:conclusionsfuturework}

In this paper, we have introduced ModelCC, a model-based tool for language specification.
ModelCC lets language designers create explicit models of the concepts a language represents, i.e. the abstract syntax model of the language (ASM).
Then, that abstract syntax can be represented in textual or graphical form, using the concrete syntax defined by a concrete syntax model (CSM).
ModelCC automates the ASM-CSM mapping by means of metadata annotations on the ASM, which let ModelCC act as a model-based parser generator.

ModelCC is not bound to particular scanning and parsing techniques, so language designers do not have to tweak their models to comply with the constraints imposed by particular parsing algorithms.
ModelCC abstracts away many details traditional language processing tools have to deal with.
It cleanly separates language specification from language processing.
Given the proper ASM-CSM mapping definition, ModelCC-generated parsers are able to automatically instantiate the ASM given an input string in the concrete syntax.

Apart from being able to deal with ambiguous languages, ModelCC also allows the declarative resolution of language ambiguities by means of constraints defined over the ASM. The current version of ModelCC also supports lexical ambiguities and custom pattern matching classes.

ModelCC incorporates a reference resolution mechanism within its parsing process.
Instead of returning abstract syntax trees, ModelCC is able to obtain abstract syntax graphs from textual inputs.
Such abstract syntax graphs are not restricted to directed acyclic graphs, since ModelCC supports the resolution of anaphoric, cataphoric, and recursive references.

The proposed model-driven language specification approach promotes the domain-driven design of language processing systems.
Its model-driven philosophy supports language evolution by improving the maintainability of such systems.
It also facilitates the reuse of language models across product lines and different applications, eliminating the duplication required by conventional tools and improving the modularity of the resulting systems.

In the future, we intend to study the possibilities ModelCC opens up in different application domains, including
traditional language processing systems (compilers and interpreters), 
domain-specific languages 
and language workbenches, 
model-driven software development tools, 
natural language processing, 
text mining, 
data integration, 
and information extraction. 

\section*{Acknowledgements}
Work partially supported by research project TIN2012-36951, ``NOESIS: Network-Oriented Exploration, Simulation, and Induction System'', funded by the Spanish Ministry of Economy and the European Regional Development Fund (FEDER).

\bibliographystyle{eptcs}
\bibliography{modelcc}

\end{document}